\documentclass[aps,pre,showpacs,preprint,superscriptaddress]{revtex4}
\usepackage{graphicx}

\begin{document}

\title{Random-energy model in random fields}

\author{Luiz O de Oliveira Filho} 
\altaffiliation[Permanent
Adress: ]{Coordena\c{c}\~ao de F\'{\i}sica, Universidade Estadual Vale
  do Acara\'u, Av. Doutor Guarani 317, 62040-730 Sobral, CE, Brazil}
\affiliation{Instituto de
  F\'{\i}sica, Universidade de S\~{a}o Paulo, Caixa Postal 66318,
  05315-970 S\~{a}o Paulo, SP, Brazil} 
\author{Francisco Alexandre da Costa} 
\affiliation{Departamento de
  F\'{\i}sica Te\'orica e Experimental, Universidade Federal do Rio
  Grande do Norte, Caixa Postal 1641, 59072-970 Natal, RN, Brazil}
\author{Carlos S O Yokoi} 
\affiliation{Instituto de F\'{\i}sica,
  Universidade de S\~{a}o Paulo, Caixa Postal 66318, 05315-970 S\~{a}o
  Paulo, SP, Brazil}

\date{\today}

\begin{abstract}
  
  The random-energy model is studied in the presence of random fields.
  The problem is solved exactly both in the microcanonical ensemble,
  without recourse to the replica method, and in the canonical
  ensemble using the replica formalism.  The phase diagrams for
  bimodal and Gaussian random fields are investigated in detail.  In
  contrast to the Gaussian case, the bimodal random field may lead to
  a tricritical point and a first-order transition. An interesting
  feature of the phase diagram is the possibility of a first-order
  transition from paramagnetic to mixed phase.

\end{abstract}


\maketitle

\section{Introduction}

Spin-glass \cite{binder86,mezard87} and random-field models
\cite{young98} have played prominent roles in the study of disordered
systems in the last few decades. Although the random-exchange and
random-field effects are  usually considered separately, it has
been argued that in proton glasses such as
Rb$_{1-x}$(NH$_4$)$_x$H$_2$PO$_4$~\cite{pirc87} it is necessary to
take into account the effect of random fields generated by the
presence of impurities. Another example where the spin-glass and
random-field effects are present simultaneously is the diluted
antiferromagnets Fe$_x$Zn$_{1-x}$F$_2$ \cite{montenegro91,vieira00}.

The effect of random fields on the well known Sherrington-Kirkpatrick
(SK) model for spin glass \cite{sherrington75} has been investigated
for Gaussian random field~\cite{soares94}, bimodal random
field~\cite{nogueira98} and trimodal random field~\cite{araujo00}.
Since the low temperature properties of the SK model is rather
difficult to work out explicitly \cite{binder86,mezard87}, it seems
worthwhile to consider a simpler spin-glass model where the effect of
random fields can be investigated thoroughly.

The random-energy model (REM)~\cite{derrida80a,derrida81} is probably
the simplest spin-glass model~\cite{gross84} retaining some important
properties of the SK model. The REM is related to the generalization
of the SK model to include interaction between every set of
$p$-spins~\cite{gross84}. In the $p \rightarrow \infty$ limit the
energies of the spin configurations become independent random
variables and the model reduces to the REM.

In this paper we investigate the effect of random fields on the REM.
The model is given as the $p \rightarrow \infty$ limit of the
Hamiltonian
\begin{equation}
\mathcal{H}=-\sum_{i_1 < \cdots < i_p} J_{i_1 \ldots i_p} S_{i_1} \cdots
S_{i_p} - J_0 \sum_{i < j} S_i S_j - \sum_i H_i S_i,
\label{Hamiltonian}
\end{equation}
where $S_i = \pm 1$ are Ising spins, $J_{i_1 \ldots i_p}$ are
independent quenched Gaussian random couplings with zero mean and
variance $p! J^2/2 N^{p-1}$, $J_0 \ge 0$ are ferromagnetic couplings
and $H_i$ are independent identically distributed quenched random
fields.  

The Hamiltonian (\ref{Hamiltonian}) for $p=2$ is the SK model in a
random field, whereas for $p \rightarrow \infty$ it reduces to the REM
model in a random field. We have solved the problem exactly by two
complementary approaches. In section 2 we employ the microcanonical
formalism~\cite{derrida81} to obtain the thermodynamic quantities
directly. In section 3 we employ the replica formalism~\cite{gross84}
to determine the spin-glass order parameters.  In section 4 we study
the phase diagram for bimodal and Gaussian distribution of random
fields. Finally, in section 5 we compare our results with the previous
studies on related models and make some concluding remarks.

\section{Microcanonical approach}

In this section we solve the model in the microcanonical ensemble
\cite{derrida81}.  Let $S = (S_1, \ldots,S_N)$ denote one of $2^N$
spin configurations or the microstates of the system. The energy of a
given microstate is given by
\begin{equation}
E_S = \mathcal{H}(S) = - \sum_{i_1< \cdots < i_p} J_{i_1 \cdots i_p} S_{i_1}
\cdots S_{i_p} + \mathcal{H}_0(S),
\end{equation}
where $\mathcal{H}_0$ denotes the part of the Hamiltonian without
random couplings. Since $E_S$ are linear combinations
of Gaussian random variables $J_{i_1 \cdots i_p}$, they are themselves
Gaussian random variables with mean
\begin{equation}
 \langle E_S \rangle  =\mathcal{H}_0(S)  = E^0_S, 
\end{equation}
and covariance
\begin{equation}
   \sigma_{S S^{\prime}} = 
  \left\langle (E_S - E^0_S)(E_{S^{\prime}} - E^0_{S^{\prime}}) 
  \right\rangle =  \frac{J^2 N}{2} \left[ q_{S      S^{\prime}}^p + O
    \left( \frac{1}{N} \right) \right],
\end{equation}
where 
\begin{equation}
  q_{S S^{\prime}} = \frac{1}{N} \sum_i S_i S^{\prime}_i,
\end{equation}
is the overlap between the microstates $S$ and $S^{\prime}$.  In the
thermodynamic limit, $N \rightarrow \infty$, the energies $E_S$ and
$E_{S^{\prime}}$ of two macroscopically distinguishable microstates
$S$ and $S^{\prime}$ become uncorrelated in the $p \rightarrow \infty$
limit,
\begin{equation}
  \sigma_{SS^{\prime}}=  \left(\frac{J^2 N}{2} \right) q_{S
    S^{\prime}}^p \longrightarrow 0, \quad \text{for $p \rightarrow
  \infty$ and  $|q_{S S^{\prime}}| < 1$}. 
\end{equation}
Thus in the $p \rightarrow \infty$ limit the energies $E_S$ become
independent Gaussian random variables. The multivariate probability
density is then the product of univariate probability densities given
by
\begin{equation}
  f_{E_S}(E) = \frac{1}{\sqrt{\pi N J^2}} \exp \left[ - \frac{(E-E^0_S)^2}{N
      J^2} \right].
\end{equation}

Let us consider a given sample, that is, a particular realization of
the random couplings $J_{i_1 \cdots i_p}$. The entropy of the sample
is given by
\begin{equation}
  S(E) = k_B \ln \Omega(E),
\end{equation}
where
\begin{equation}
  \Omega(E) = \sum_S \delta  (E - E_S)
\end{equation}
is the density of states. The average density of states is
\begin{equation}
  \langle \Omega(E) \rangle = \sum_S \langle  \delta
  (E - E_S)  \rangle   = \sum_S f_{E_S}(E).
\end{equation}
Due to the statistical independence of $E_S$ the fluctuations around
this average is of order $\langle \Omega(E) \rangle^{-1/2}$, and
thus completely negligible~\cite{derrida81}.

We can rewrite the average density of states in the form
\begin{equation}
  \langle \Omega(E) \rangle   =\frac{1}{\sqrt{\pi N J^2}}
  \int_{-\infty}^{\infty} d E_0 \exp \left[ - \frac{(E-E_0)^2}{N
      J^2} \right] \sum_S \delta(E_0 - E^0_S).
\end{equation}
We recognize 
\begin{equation}
  \Omega_0(E) = \sum_S \delta  (E_0 - E^0_S)
\end{equation}
as the density of states of the system described by the Hamiltonian
$\mathcal{H}_0$. Therefore
\begin{equation}
 \langle \Omega(E) \rangle = \frac{1}{\sqrt{\pi N J^2}}
 \int_{-\infty}^{\infty} d E_0 \exp \left[ -\frac{(E - E_0)^2}{NJ^2} +
   \frac{S_0(E_0)}{k_B}  \right],
\end{equation}
where
\begin{equation}
  S_0(E_0) = k_B \ln \Omega_0(E_0),
\end{equation}
is the entropy of the system characterized by the Hamiltonian 
$\mathcal{H}_0$. In the thermodynamic limit, $N \rightarrow \infty$,
we have
\begin{equation}
\ln  \langle \Omega(E) \rangle  = \max_{E_0}  \left[ -\frac{(E - E_0)^2}{NJ^2} +
   \frac{S_0(E_0)}{k_B}  \right]. 
\end{equation}
$E_0$ is determined by
\begin{equation}
  \frac{1}{T_0(E_0)} = \frac{ \partial     S_0(E_0)}{\partial E_0} =  -
  \frac{2k_B(E - E_0)}{NJ^2},
\end{equation}
where $T_0(E_0)$ is by definition the temperature of the system
described by the Hamiltonian $\mathcal{H}_0$.

For energies $E$ such that $\ln \langle \Omega(E) \rangle
> 0$  the average density of states is very large and the fluctuation
is negligible. Thus we have with probability 1,
\begin{equation}
  S(E)= k_B \ln  \langle \Omega(E) \rangle =  -\frac{k_B(E -
    E_0)^2}{NJ^2} + S_0(E_0). 
\end{equation}
For energies $E$  such that $\ln \langle \Omega(E) \rangle
< 0$, the average density of states is very small. Thus with 
probability 1 there are no samples with this energy.

The temperature of the system is given by
\begin{equation}
\frac{1}{T(E)}= \frac{ \partial     S(E)}{\partial E} =   -\frac{2 k_B(E -
  E_0)}{NJ^2},  
\end{equation}
which coincides with the temperature of the system described by the
Hamiltonian $\mathcal{H}_0$,
\begin{equation}
  T(E) = T_0 (E_0).
\end{equation}
Therefore the energy of the system as a function of temperature is
given by
\begin{equation}
  E(T) = E_0(T) - \frac{N J^2}{2 k_B T},
\end{equation}
where $E_0(T)$ is the energy of the system characterized by the
Hamiltonian $\mathcal{H}_0$. The entropy as a function of the
temperature is 
\begin{equation}
  S(T) = S_0(T)- \frac{N J^2}{4 k_B T^2}.
\end{equation}
These results are valid above a critical temperature $T_c$ determined by
\begin{equation}
  S(T_c) = S_0(T_c)- \frac{N J^2}{4 k_B T_c^2} = 0.
\end{equation}
Below this temperature the system is frozen in its ground state.

These results are valid for any Hamiltonian $\mathcal{H}_0$. We now
particularize for the case where the Hamiltonian $\mathcal{H}_0$
describes the Ising model with infinite range ferromagnetic
interactions in a random field~\cite{schneider77,aharony78,salinas85},
\begin{equation}
  \mathcal{H}_0= - \frac{J_0}{N} \sum_{i < j} S_i S_j - \sum_i H_i S_i
  =   - \frac{J_0}{2N} \left( \sum_i S_i \right)^2 - \sum_i
  H_i S_i,
\end{equation}
where in the last passage we have dropped the term $J_0/2$ that is
negligible in the thermodynamic limit. The quadratic term can be
linearized using the identity
\begin{equation}
  e^{\lambda a^2/2} = \sqrt{\frac{\lambda}{2\pi}}
  \int_{-\infty}^{\infty} dx e^{-\lambda x^2/2 + \lambda a x},
\label{gaussian_identity}
\end{equation}
and the partition function will be given by
\begin{equation}
  Z_0 = \sum_S e^{-\beta \mathcal{H}_0} = \sqrt{\frac{\beta J_0 N}{2
      \pi}}  \int_{-\infty}^{\infty} dm 
    \exp \left\{ N \left[ - \frac{1}{2}\beta J_0 m^2 + \frac{1}{N} \sum_i \ln
      2 \cosh \beta(J_0 m +H_i) \right] \right\}.
\end{equation}
In the thermodynamic limit, $N \rightarrow \infty$, the Laplace method
gives
\begin{equation}
 \ln Z_0 = N \max_m  \left[ - \frac{1}{2}\beta J_0 m^2 + \left\langle
          \ln       2 \cosh \beta(J_0 m +H) \right\rangle \right],
\end{equation}
where we have used the law of large numbers to write
\begin{equation}
   \frac{1}{N} \sum_i \ln       2 \cosh \beta(J_0 m +H_i) =
   \left\langle  \ln       2 \cosh \beta(J_0 m +H) \right\rangle,
\end{equation}
where $\langle \cdots \rangle$ denotes the expectation value with
respect to the random fields $H$.
 Thus the free energy is given by
\begin{equation}
F_0 = -\beta^{-1} \ln Z_0 = N \left[ \frac{1}{2} J_0 m^2 - \frac{1}{\beta}
\left\langle \ln 2 \cosh \beta(J_0 m +H) \right\rangle \right],
\end{equation}
where the magnetization $m$ is determined by the equation 
\begin{equation}
  m = \left\langle \tanh \beta(J_0 m +H) \right\rangle.
\end{equation}
The internal energy $E_0(T)$ and the entropy $S_0(T)$ follow from
usual thermodynamic relations.

Applying the general results obtained previously for the system
described by the full Hamiltonian (\ref{Hamiltonian}) in the $p
\rightarrow \infty$ limit,  we obtain for the internal energy
\begin{equation}
  \frac{E}{N}  =   - \frac{\beta J^2}{2}
    - \frac{1}{2} J_0 m^2  -  \left\langle H \tanh \beta(J_0 m +H)
  \right\rangle,
\end{equation}
and for the entropy
\begin{equation}
  \frac{S}{Nk_B} =
  -\frac{(\beta        J)^2}{4}   - \beta J_0 m^2 - \beta
    \left\langle H \tanh \beta(J_0 m +H)   \right\rangle 
     +     \left\langle \ln 2 \cosh \beta(J_0 m +H) \right\rangle.
\end{equation}
These results are valid for $\beta < \beta_c$ where $\beta_c$ is
determined by
\begin{equation}
  S(\beta_c) =  - \frac{(\beta_c J)^2}{4} -  \beta_c \langle  (H+J_0m)\tanh \beta_c
  (H+J_0m) \rangle  + \langle \ln 2 \cosh  \beta_c 
  (H+J_0m)  \rangle  = 0
\label{entropy0}
\end{equation}
and
\begin{equation}
  m = \left\langle \tanh \beta_c(J_0 m +H) \right\rangle.
\end{equation}
For $\beta > \beta_c$ the system is frozen in its ground
state. Therefore 
\begin{equation}
E(\beta) = E(\beta_c), \qquad  S(\beta) = 0.
\end{equation}

\section{Replica approach}

In this section we solve the model in the canonical ensemble
\cite{gross84}. We use the replica identity for the free energy
\begin{equation}
  -\beta F = \langle \ln Z \rangle =
  \lim_{n \rightarrow 0} \frac{\langle Z^n  \rangle -1}{n},
\end{equation}
to perform the average over the random couplings $J_{i_1 i_2 \cdots
  i_p}$. To evaluate $\langle Z^n  \rangle$ we introduce $n$ replicas
of the system $\alpha=1,2,\ldots,n$,
\begin{equation}
  \left\langle Z^n \right\rangle = \text{Tr}\, \left\langle e^{-\beta
      \sum_{\alpha=1}^n \mathcal{H}(S^{\alpha})} \right\rangle =
  \text{Tr}\, e^{-\beta \mathcal{H}_{\text{eff}}}, 
\end{equation}
where $\mathcal{H}_{\text{eff}}$ denotes the effective Hamiltonian
that results after taking the average over random couplings,
\begin{equation}
-\beta \mathcal{H}_{\text{eff}} = \frac{N(\beta J)^2}{2} \left[
  \sum_{\alpha < \beta} \left( \frac{1}{N} \sum_i S_i^{\alpha}
    S_i^{\beta}\right)^p + \frac{n}{2} \right] + \frac{N \beta J_0}{2}
\sum_{\alpha} \left(\frac{1}{N} \sum_i S_i^{\alpha} \right)^2 + \beta
\sum_i H_i \sum_{\alpha} S_i^{\alpha}. 
\end{equation}
We have dropped terms that vanish in the thermodynamic limit, $N
\rightarrow \infty$.  The nonlinear terms can be linearized with the
help of the asymptotic relation
\begin{equation}
  e^{N \lambda f(a)} \sim  \sqrt{\frac{ N \lambda f^{\prime \prime}
      (a)}{2 \pi } }
  \int_{-\infty}^{\infty}  dx e^{N 
    \lambda \left[     f(x)  - f^{\prime}(x) (x-a) \right] },
\end{equation}
which can be proved for $\lambda f^{\prime \prime}(a) > 0$ and $N
\rightarrow \infty$ applying the Laplace method. In particular for
$f(x)=x^2/2$ the asymptotic relation reduces to the identity
(\ref{gaussian_identity}). Omitting the factors that do not contribute
to the free energy in the thermodynamic limit, $N \rightarrow \infty$,
we arrive at
\begin{equation}
   \langle Z^n \rangle \sim \int \prod_{\alpha < \beta}  
  d q_{\alpha \beta} \int \prod_{\alpha} d m_{\alpha} e^{- \beta
    F_n(q_{\alpha \beta}, m_{\alpha}) },
\end{equation}
where
\begin{eqnarray}
  \frac{F_n}{N} &=&  -\frac{1}{4} \beta J^2
  n + \frac{1}{2} \beta J^2(p-1) 
  \sum_{\alpha < \beta}q_{\alpha \beta}^p + \frac{1}{2} 
  J_0 \sum_{\alpha} m_{\alpha}^2 \nonumber \\&& - \beta^{-1}
\frac{1}{N}  \sum_i \ln   \text{Tr}\,  \exp \left[
    \frac{1}{2} (\beta J)^2 p       \sum_{\alpha < \beta} q_{\alpha
      \beta}^{p-1}  S^{\alpha}   S^{\beta}  
      + \beta \sum_{\alpha} (H_i + J_0 m_{\alpha})  S^{\alpha}
      \right].
\end{eqnarray}
In the $N \rightarrow \infty$ limit we use the law of large numbers to
write the last term as an expectation value over the random-field
distribution and use Laplace method to evaluate the integral. The free
energy is then given by the stationary value of the functional
\begin{eqnarray}
  \frac{F}{N} &=& -\frac{1}{4}\beta J^2 +\lim_{n
    \rightarrow 0} \frac{1}{n}  \Bigg\{
    \frac{1}{2}\beta J^2(p-1) \sum_{\alpha < \beta} q_{\alpha \beta}^p +
    \frac{J_0}{2} \sum_{\alpha} m_{\alpha}^2 \nonumber \\&&  - \beta^{-1}
    \left\langle \ln \text{Tr}\, 
    \exp \Bigg[ \frac{1}{2}(\beta J)^2 p
  \sum_{\alpha < \beta} q_{\alpha \beta}^{p-1} S^{\alpha}
  S^{\beta} 
  + \beta \sum_{\alpha} (H + J_0 m_{\alpha}) S^{\alpha} \Bigg]
\right\rangle \Bigg\}, 
\end{eqnarray}
where  $\langle \cdots \rangle$ denotes the expectation value with
respect to the random field $H$.

To compute the free energy we assume
\begin{equation}
  m_{\alpha} = m,
\end{equation}
to be independent of replica indices, and parameterize $q_{\alpha
  \beta}$ following the Parisi's $K$-step replica-symmetry-breaking
Ansatz~\cite{parisi80}. In the $n \rightarrow 0$ limit the free energy
functional becomes a function of the magnetization $m$ and the
parameters
\begin{equation}
  0  \le q_0 \le q_1 \le \cdots \le q_{K-1} \le q_K \le 1,
\end{equation}
\begin{equation}
  0 = m_0 \le m_1 \le \cdots m_K \le m_{K+1} = 1,
\end{equation}
and is given by
\begin{eqnarray}
  \frac{F}{N} &=& -\frac{\beta J^2}{4} \left[ 1 + (p-1)
    \sum_{i=0}^K (m_{i+1} - m_i)q_i^p - p q_K^{p-1} \right] +
  \frac{J_0}{2} m^2 \nonumber \\ &&  - \int_{-\infty}^{\infty}
  dy \left\langle G_{\sigma_0^2}(y-H-J_0m) \right\rangle g_0(y),
\end{eqnarray}
where $g_0(y)$ is given recursively by
\begin{equation}
  g_{i-1}(y)= \frac{1}{\beta m_i} \ln \Bigg\{ \int_{-\infty}^{\infty} d
  y^{\prime} G_{\sigma_i^2}(y^{\prime} 
  -y) \exp \big[ \beta m_i g_i(y^{\prime}) \big] \Bigg\},
\end{equation}
for $i=1,\ldots,K$ with the initial condition
\begin{equation}
  g_K(y) = \frac{1}{\beta} \ln (2 \cosh \beta y). 
\end{equation}
$G_{\sigma^2}(y)$ denotes the  Gaussian distribution function 
\begin{equation}
  G_{\sigma^2}(y) = \frac{1}{J \sigma\sqrt{2 \pi }} \exp \left( -\frac{y^2}{2 J^2 \sigma^2}  \right),
\end{equation}
where the variances $\sigma_i^2$ are given by
\begin{equation}
  \sigma_0^2 =\frac{p}{2}q_0^{p-1}, \qquad
  \sigma_i^2=\frac{p}{2}(q_i^{p-1}-q_{i-1}^{p-1}) \qquad \text{for}
  \qquad i=1,\ldots,K.
\end{equation}

We first assume that all the $q$'s are less than one, $0 \le q_0 \le
\ldots \le q_{K-1} \le q_K < 1$. Then $\sigma_i^2 \rightarrow 0$ when
$p \rightarrow \infty$ for $i=0,\ldots,K$. Using the expansion
\begin{equation}
  \int_{-\infty}^{\infty} d y^{\prime} G_{\sigma^2}(y^{\prime}-y)f(y^{\prime})
  =  \exp \left(\frac{J^2 \sigma^2}{2} \frac{d^2}{d y^2} \right)
  f(y) = 1 + \frac{J^2 \sigma^2}{2} f^{\prime\prime}(y) + O(\sigma^4),
\label{expansion}
\end{equation}
we obtain
\begin{eqnarray}
  \frac{F}{N}  &=&  -\frac{\beta J^2}{4} \left[ 1 + (p-1)
    \sum_{i=0}^K (m_{i+1} - m_i)q_i^p - p q_K^{p-1} \right] +
  \frac{J_0}{2} m^2 -  \frac{1}{\beta} \langle \ln 2 \cosh \beta (H +
  J_0 m) \rangle   \nonumber \\ &&  - \sum_{i=0}^K  \frac{\beta (J
    \sigma_i)^2}{2} \left[ 1 
  -(1-m_i) \langle \tanh^2 \beta  (H + J_0 m) \rangle \right] +
      O(\sigma_0^4,\ldots,\sigma_K^4,\sigma_0^2
      \sigma_1^2,\ldots,\sigma_0^2 \sigma_K^2).\nonumber  \\
\end{eqnarray}
Stationarity of the free energy with respect to the variational
parameters gives, in the limit $p \rightarrow \infty$, 
\begin{equation}
  m = \langle \tanh  \beta (H + J_0 m ) \rangle,
\end{equation}
and 
\begin{equation}
  q_0= q_1= \cdots = q_K = \langle \tanh^2  \beta (H + J_0 m ) \rangle.
\end{equation}
Thus we arrived at the replica-symmetric solution where all the $q$'s
are identical. The free energy in the $p \rightarrow \infty$ limit is
given by
\begin{equation}
  \frac{F}{N} = -\frac{\beta J^2}{4} +  \frac{J_0}{2} m^2 - \frac{1}{\beta}
  \langle \ln  2 \cosh \beta (H + J_0 m ) \rangle.
\end{equation}
The entropy is
\begin{equation}
  \frac{S}{Nk_B} =   -\frac{(\beta        J)^2}{4}  - \beta
    \left\langle (J_0 m + H) \tanh \beta(J_0 m +H)   \right\rangle 
     +     \left\langle \ln 2 \cosh \beta(J_0 m +H) \right\rangle.
\end{equation}
This solution corresponds precisely to the high temperature solution
found in the microcanonical approach. Since the entropy becomes
negative at low temperatures, it is necessary to consider a different
solution for low temperatures.

We therefore assume that $0 \le q_0 \le \ldots \le q_{K-1} < q_K =1$.
Then $\sigma_i^2 \rightarrow 0$ for $i=0,\ldots,K-1$ and $\sigma_K
\rightarrow \infty$ in the limit $p \rightarrow \infty$. A simple
calculation yields,
\begin{equation}
  g_{K-1}(y)=\frac{1}{\beta m_K} \ln  (2 \cosh \beta m_K y) + 
  \frac{1}{2}\beta m_K(J \sigma_K)^2 +  O( e^{-(\beta J \sigma_K m_K)^2/2}\sigma_K^{-1}).
\end{equation}
The error is exponentially small and may be safely ignored. The rest of
calculation proceeds as before using the expansion (\ref{expansion}) and
we arrive at
\begin{eqnarray}
  \frac{F}{N} &=&  -\frac{\beta J^2}{4} \left[ 1 + (p-1)
    \sum_{i=0}^K (m_{i+1} - m_i)q_i^p - p q_K^{p-1} \right] +
  \frac{J_0}{2} m^2  -  \frac{1}{2}\beta m_K(J \sigma_K)^2 \nonumber
  \\ && 
  -\sum_{i=0}^{K-1}
  \frac{\beta (J \sigma_i)^2}{2}       \big[ m_K \langle \text{sech}^2 \beta m_K
  (H+J_0m) \rangle       + m_i  \langle \tanh^2 \beta m_K (H+J_0m)
  \rangle   \big] \nonumber \\ &&     -  \frac{1}{\beta m_K} \langle \ln 2 \cosh  \beta
  m_K (H+J_0m) \rangle +       O(\sigma_0^4,\ldots,\sigma_{K-1}^4,\sigma_0^2
      \sigma_1^2,\ldots,\sigma_0^2 \sigma_{K-1}^2).
\end{eqnarray}
Stationarity with respect to the variational parameters gives, in the
limit $p \rightarrow \infty$, 
\begin{equation}
  m = \langle \tanh  \beta m_K (H + J_0 m ) \rangle,
\label{m}
\end{equation}
\begin{equation}
  q_0=q_1=\cdots=q_{K-1} = \langle \tanh^2  \beta m_K (H + J_0 m )
  \rangle, \qquad q_K =1, 
\end{equation}
consistent with initial assumption $q_K=1$, and
\begin{equation}
  \frac{(\beta J)^2}{4} m_K^2 = \langle \ln 2 \cosh  \beta m_K
  (H+J_0m)  \rangle  -  \beta m_K \langle  (H+J_0m)\tanh \beta m_K
  (H+J_0m) \rangle.
\label{m_K}
\end{equation}
These results are the same for all $K \ge 1$, showing that no other
solutions are possible beyond one-step replica symmetry breaking.
The free energy in the limit $p \rightarrow \infty$ is given by
\begin{equation}
  \frac{F}{N} = -\frac{\beta J^2}{4} m_K +  \frac{J_0}{2} m^2 - \frac{1}{\beta
    m_K} \langle \ln   2 \cosh \beta m_K(H + J_0 m ) \rangle.
\end{equation}
The entropy is
\begin{equation}
  \frac{S}{Nk_B} =   -\frac{(\beta J)^2}{4}m_K -  \beta 
    \left\langle (J_0 m +H)  \tanh \beta m_K (J_0 m +H)   \right\rangle 
     +     \frac{1}{m_K} \left\langle \ln 2 \cosh \beta m_K (J_0 m +H) \right\rangle.
\end{equation}
Taking into account the self-consistency equation (\ref{m_K}) we find
that the entropy vanishes identically. Thus this solution corresponds
to the frozen phase found in the microcanonical approach.

The self-consistency equations (\ref{m}) and (\ref{m_K}) imply that 
$\beta m_K$ is independent of temperature. Since this solution is
acceptable only for $m_K \le 1$, we have
\begin{equation}
  \beta m_K = \beta_c, 
\end{equation}
where $\beta_c$ is found from the equations
\begin{equation}
  \frac{(\beta_c J)^2}{4} = \langle \ln 2 \cosh  \beta_c 
  (H+J_0m)  \rangle  -  \beta_c \langle  (H+J_0m)\tanh \beta_c
  (H+J_0m) \rangle,
\end{equation}
and
\begin{equation}
  m = \langle \tanh  \beta_c (H + J_0 m ) \rangle.
\end{equation}
Thus we see that $\beta_c$ corresponds precisely to the critical
temperature for the transition to the frozen phase found in the
microcanonical approach. The Parisi order parameter function $q(x)$
\cite{parisi80} has two flat portions $q_0=m^2$ and $q_K=1$, with a
discontinuous jump at $x=m_K=T/T_c$,
\begin{equation}
  q(x)=m^2 \theta \left(\frac{T}{T_c} - x \right) + \theta \left(x -
    \frac{T}{T_c}\right).
\label{q(x)}
\end{equation}
The overlap distribution function $P(q)$~\cite{parisi83} is given by
\begin{equation}
  P(q)=\frac{T}{T_c} \delta(q-m^2) + \left(1 - \frac{T}{T_c} \right)
  \delta(q-1).
\end{equation}
Thus the frozen phase is indeed a spin-glass phase with many pure
states having minimal overlap between them and maximal self-ovelap
\cite{gross84}. 

\section{Phase diagrams}

The phase diagrams for the model defined by the Hamiltonian
(\ref{Hamiltonian}) in the $p \rightarrow \infty$ limit  were
determined for two distributions of random fields which are often
considered in the literature: The discrete bimodal distribution~\cite{aharony78}
\begin{equation}
  P(H)=\frac{1}{2} \delta(H - H_0 + \sigma) + \frac{1}{2} \delta(H -
  H_0 - \sigma),
\end{equation}
and the continuous Gaussian distribution function~\cite{schneider77}
\begin{equation}
  P(H)=\frac{1}{\sqrt{2 \pi} \sigma} e^{-(H - H_0)^2/2 \sigma^2}.
\end{equation}
In both cases the means and variances are $H_0$ and $\sigma^2$,
respectively.

\begin{figure}
 \includegraphics{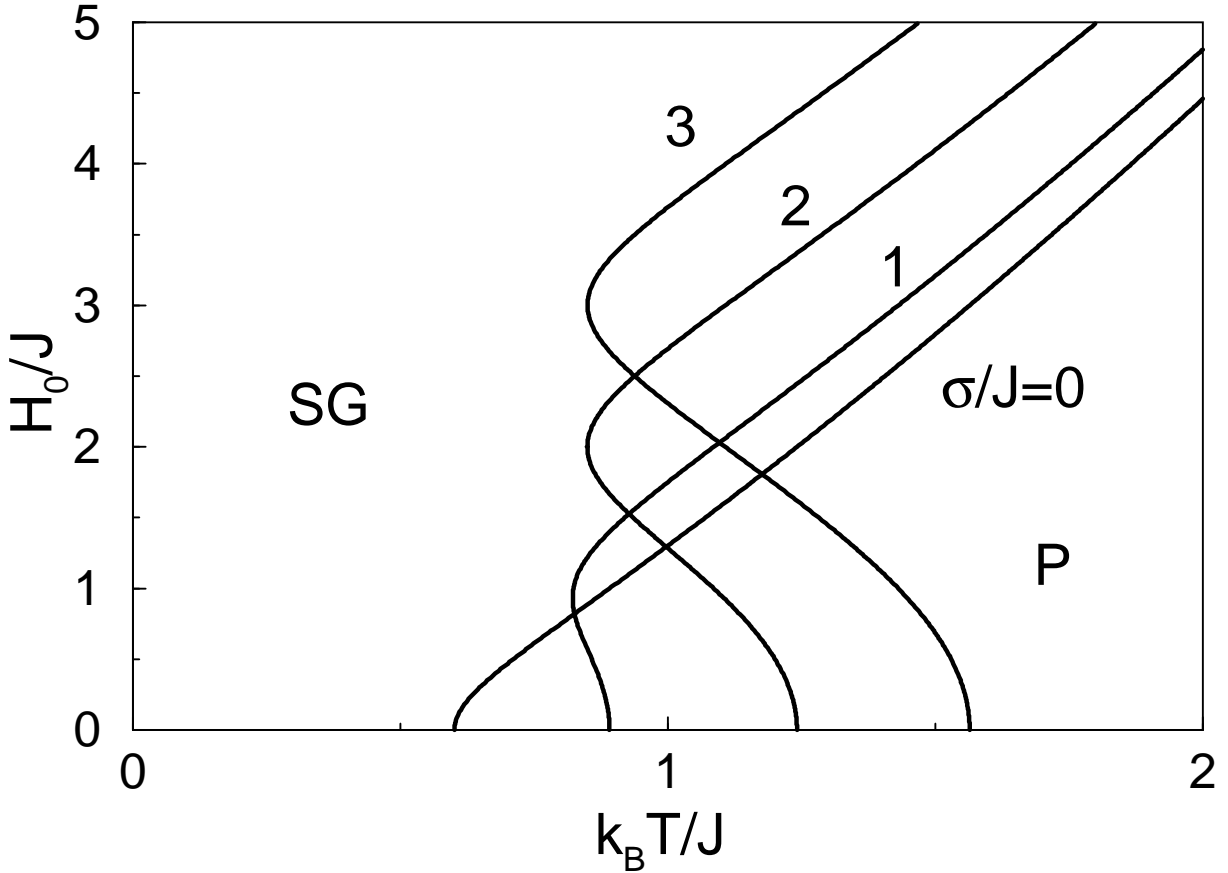}
 \caption{\label{txh0_b} The $J_0=0$ phase diagram for bimodal random
   field for various values of  $\sigma$.}
\end{figure}

\begin{figure}
 \includegraphics{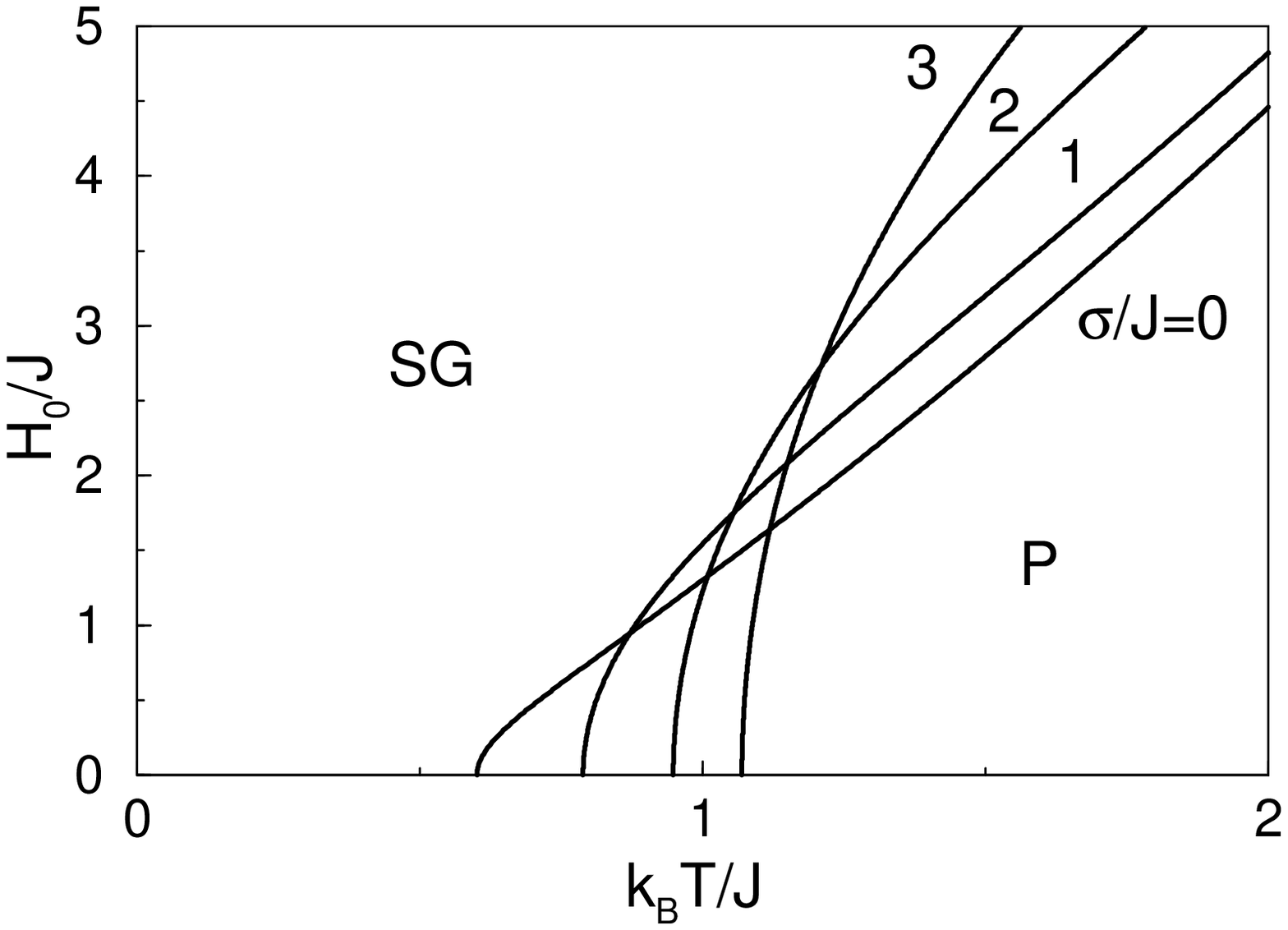}
 \caption{\label{txh0_g} The $J_0=0$ phase diagram for Gaussian random
   field for various values of  $\sigma$.}
\end{figure}

The $H_0 \times T$ phase diagrams in the absence of ferromagnetic
interactions ($J_0=0$) and various values of the standard deviation
$\sigma$ are shown in Fig.~\ref{txh0_b} and Fig.~\ref{txh0_g} for
bimodal and Gaussian distributions, respectively. Notice that the case
$\sigma=0$ reduces to the REM in a uniform field~\cite{derrida81}.
There are two phases: replica-symmetric paramagnetic phase (P) and a
frozen spin-glass (SG) phase with one-step replica symmetry breaking.
The transitions are determined by equation (\ref{entropy0}) for
$J_0=0$,
\begin{equation}
  - \frac{(\beta_c        J)^2}{4}   - 
  \beta_c     \left\langle H \tanh \beta_c H  \right\rangle 
     +     \left\langle \ln 2 \cosh \beta_c H \right\rangle = 0.
\label{P-SG}
\end{equation}
These transitions are second order in the thermodynamic sense, but the
Parisi order parameter (\ref{q(x)}) changes discontinuously at the
transition.  For both distributions the main effect of the disorder
$\sigma > 0$ is to increase the transition temperature for small $H_0$
and depress it for large $H_0$ when compared to the case $\sigma=0$.
This effect is most pronounced for the bimodal distribution, shown in
Fig.~\ref{txh0_b}. For $\sigma > 0$, as $H_0$ departs from zero, the
transition temperature initially decreases, reaches a minimum, and
then starts to increase as a function of $H_0$. For $\sigma \gg J$ the
minimum of the transition temperature occurs for $H_0 \simeq \sigma$,
as can be derived from Eq.~\ref{P-SG}. For the gaussian case, on the
contrary, the transition temperature increases monotonically as a
function of $H_0$, as shown in Fig.~\ref{txh0_g}.

\begin{figure}
 \includegraphics{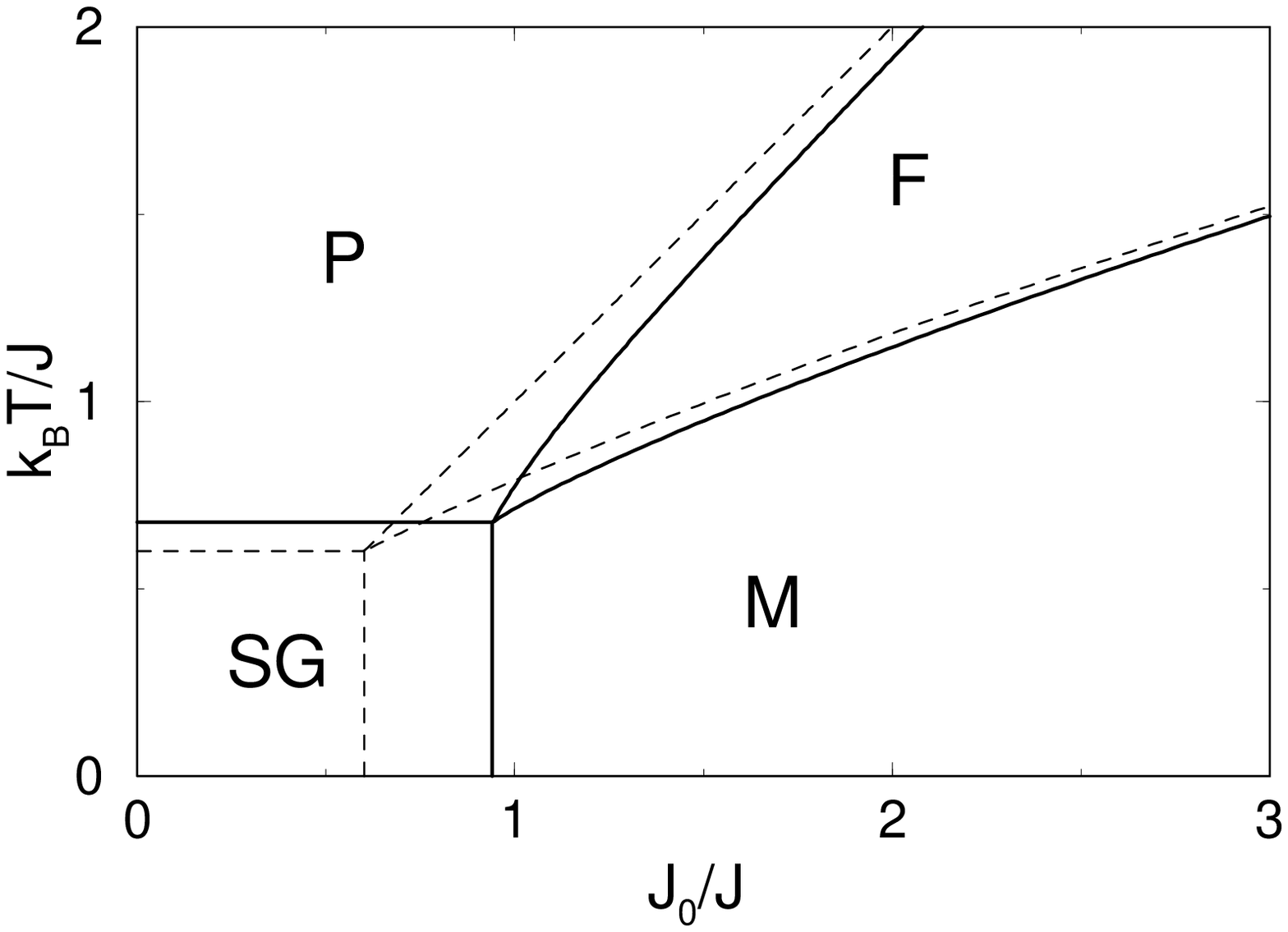}
 \caption{\label{txj0_b_04}The phase diagram for symmetric bimodal
 random field. The results for $\sigma/J =0.4$ are shown by the solid
 curves. For comparison the results in the absence of random fields
 are shown by the dashed curves.}
\end{figure}

\begin{figure}
 \includegraphics{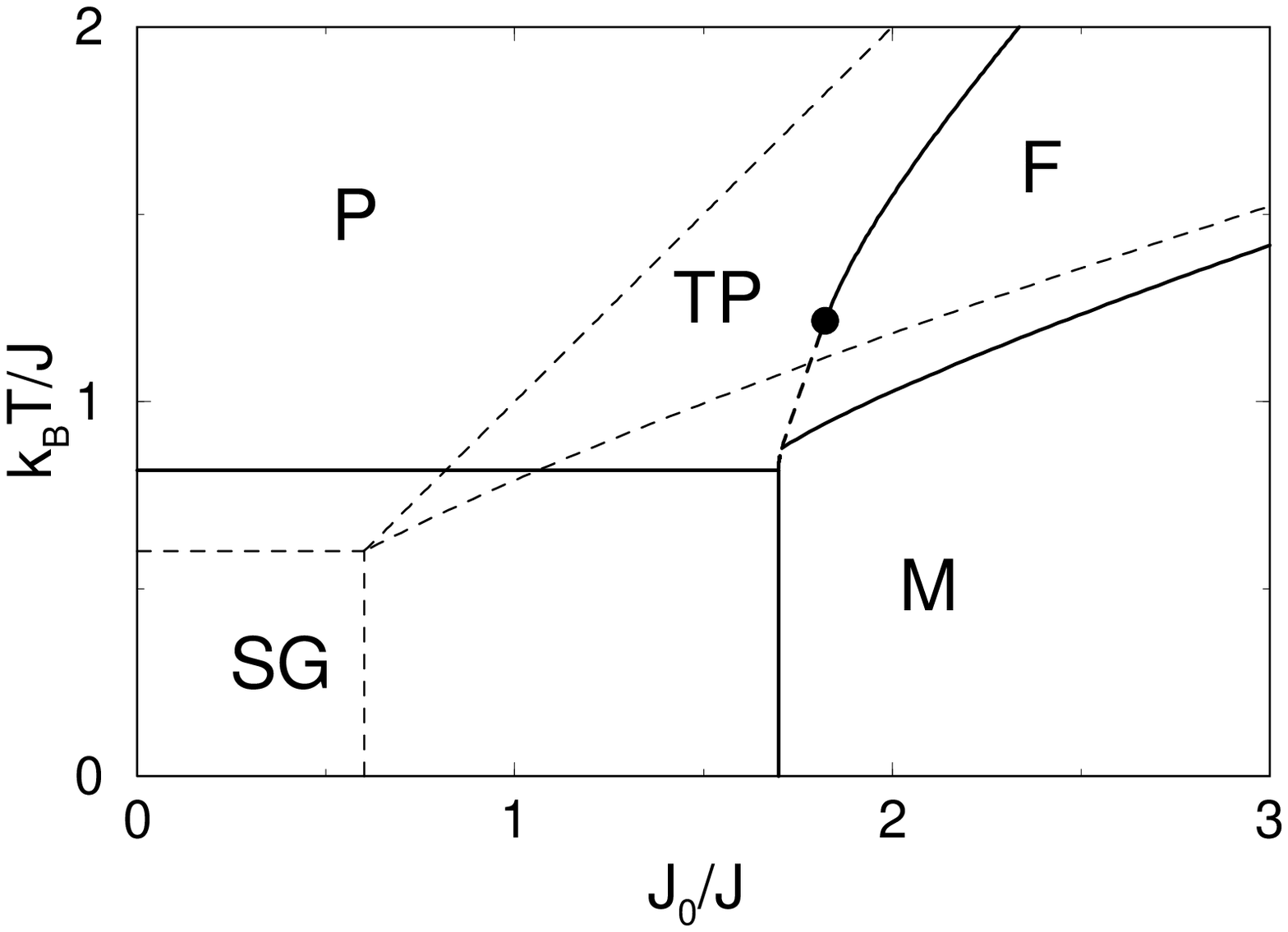}
 \caption{\label{txj0_b_08}The phase diagram for symmetric bimodal
   random field. The results for $\sigma/J =0.8$ are shown by the solid
   curves (second-order transition) and thick dashed curves (first-order
   transition). For comparison the results in the absence of random
   fields are shown by the thin dashed curves.}
\end{figure}

\begin{figure}
 \includegraphics{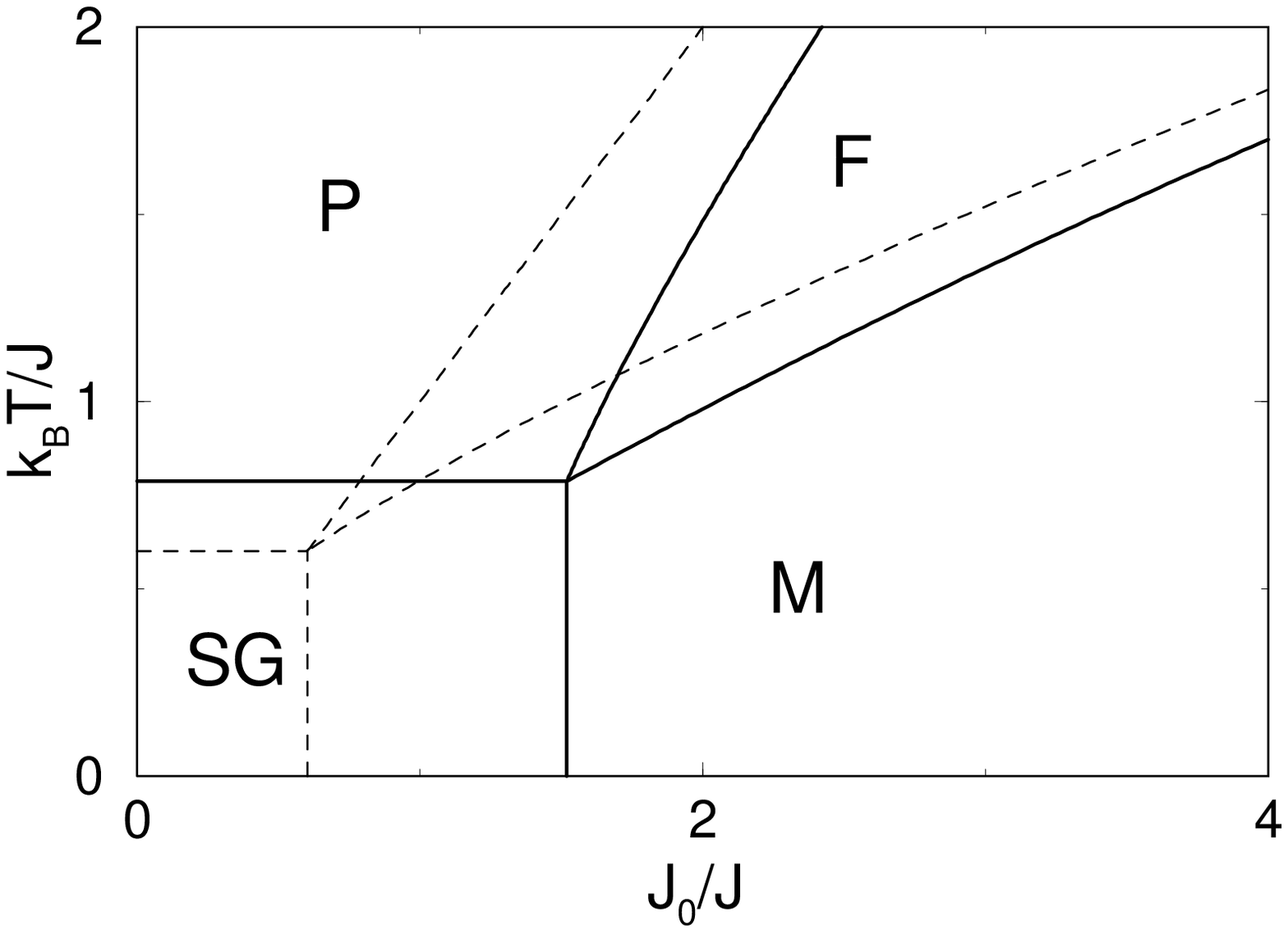}
 \caption{\label{txj0_g_1}The phase diagram for symmetric Gaussian
 random field. The results for $\sigma/J =1$ are shown by the solid
 curves. For comparison the results in the absence of random fields
 are shown by the dashed curves.}
\end{figure}

The $T \times J_0$ phase diagrams for symmetric random-field
distributions ($H_0=0$) are shown in Figs.~\ref{txj0_b_04} to
\ref{txj0_g_1}. For comparison the phase diagrams for the case
$\sigma=0$, corresponding to the REM with ferromagnetic interactions,
are shown by thin dashed lines.  There are four phases:
replica-symmetric paramagnetic (P) and ferromagnetic (F) phases, and
frozen spin-glass (SG) and mixed (M) phases with one-step replica
symmetry breaking. Unlike the SG phase, in the M phase there is a
non-zero magnetization ($m \ne 0$).

The transition from P to SG phase is determined by equation
(\ref{entropy0}) for $m=0$ which is identical to
equation~(\ref{P-SG}). Since there is no dependence on $J_0$, it
represents an horizontal line in the $T \times J_0$ phase diagram.
This transition is second order in the thermodynamic sense, but the
Parisi order parameter (\ref{q(x)}) changes discontinuously at the
transition.

A second-order transition from the P to F phase can be determined by
expanding the equation of state
\begin{equation}
  m = \left\langle \tanh \beta (H + J_0 m)  \right\rangle,
\end{equation}
in powers of the magnetization $m$~\cite{aharony78}. For symmetric
distribution of the random fields ($H_0=0$) we find
\begin{equation}
  m = a m - b m^3 - c m^5 -  \cdots,
\end{equation}
where
\begin{eqnarray}
a&=&\beta J_0 \left(1- \left\langle \tanh^2 \beta H \right\rangle \right),\\
b&=&\frac{1}{3} (\beta J_0)^3 \left(1 - 4 \left\langle \tanh^2 \beta H
  \right\rangle +3 \left\langle\tanh^4 \beta H \right\rangle \right), \\ 
c&=&-\frac{1}{15} (\beta J_0)^5  \left(2 - 17 \left\langle \tanh^2 \beta H
  \right\rangle +30 \left\langle\tanh^4 \beta H
  \right\rangle-15\left\langle\tanh^6 \beta H \right\rangle  \right).
\end{eqnarray}
There is a second-order transition from P to F phase for $a=1$ and $b
>0$. For $b=0$ there is a tricritical point, and for $b<0$ the
transition is first order and can only be determined numerically by
equating the free energies of both phases.

For the bimodal distribution of random fields, the conditions $a=b=0$
gives
\begin{equation}
    \beta  \sigma = \tanh^{-1} \left(\frac{1}{\sqrt{3}}\right) =
  \frac{1}{2} \ln (2 + \sqrt{3}),
\end{equation}
which determines the location of the tricritical point. 
This tricritical point occurs above the freezing transition
(\ref{P-SG}) only if the standard deviation $\sigma$ is greater than
the threshold value
\begin{equation}
  \frac{\sigma_c}{J} =  \frac{1}{4}\ln (2 + \sqrt{3})  \left[ \ln
   \left( 3 + \sqrt{3} \right)-\frac{1}{6}\left(3 + \sqrt{3}\right) 
 \ln \left( 2 + \sqrt{3} \right) \right]^{-1/2} = 0.4584695507 \ldots.
\end{equation}
Thus the phase diagrams for the bimodal distribution are qualitatively
different depending on the value of standard deviation $\sigma$. 

For $\sigma < \sigma_c$ the phase diagrams do not differ qualitatively
from the case without random fields, as shown in Fig.~\ref{txj0_b_04}.
All the transition lines are second order in the thermodynamic sense.
However, across the F-M and SG-M transitions the Parisi order parameter
(\ref{q(x)}) changes discontinuously.

For $\sigma > \sigma_c$ the phase diagrams changes qualitatively
compared to the case without random fields.  The second-order P-F
transition line ends at a tricritical point, below which the
transition is of first order shown by thick dashed line in
Fig.~\ref{txj0_b_08}.  This line was determined by equating the free
energies of the two neighboring phases. The transition is of first
order in the thermodynamic sense as well as in the discontinuity of
the Parisi order parameter (\ref{q(x)}) across the transition.

For the Gaussian distribution one always has $b>0$ when $a=1$. Thus
the P-F transition is always of second order and the phase diagrams
and the nature of the transitions do not differ qualitatively from the
case of bimodal distribution for $\sigma < \sigma_c$, as shown in
Fig.~\ref{txj0_g_1}.

\section{Discussion}

We solved exactly the REM in a random field in both microcanonical and
replica approaches. We investigated in detail the phase diagrams for
bimodal and Gaussian random-field distributions with mean $H_0$ and
variance $\sigma^2$. The Gaussian random fields do not change the
phase diagrams qualitatively. The bimodal random fields, on the
contrary, changes the $H_0=0$ phase diagram qualitatively for
sufficiently large $\sigma$ by leading to a tricritical point and a
first-order transition at low temperatures. The same conclusions were
reached in the replica-symmetric study of the SK model in a Gaussian
random field~\cite{soares94} and bimodal random fields
\cite{nogueira98}.

For a ferromagnet it has been shown that a random-field distribution
with a minimum at zero field leads to a tricritical point and
first-order transition~\cite{aharony78,salinas85}. It is likely that
this property remains true even when random interactions are included,
although at low temperatures the emergence of the tricritical point
may be forestalled by the spin-glass phase~\cite{araujo00}.

Our results can shed useful light on the nature of phase diagram of
more sophisticated spin-glass models with random
fields~\cite{soares94,nogueira98,araujo00}. One interesting feature of
the phase diagram with tricritical point is the possibility of a
first-order transition from paramagnetic to mixed phase.

\begin{acknowledgments}
  This work was supported in part by the Brazilian Government Agencies
  CAPES and CNPq.
\end{acknowledgments}

\end{document}